\documentclass[a4paper,
               biblatex,     
               ]{jacow}
\usepackage{pdfpages,multirow,ragged2e} %
\makeatletter%
	\ifboolexpr{bool{xetex}}
	 {\renewcommand{\Gin@extensions}{.pdf,%
	                    .png,.jpg,.bmp,.pict,.tif,.psd,.mac,.sga,.tga,.gif,%
	                    .eps,.ps,%
	                    }}{}
\makeatother

\ifboolexpr{bool{xetex} or bool{luatex}} 
 {}                                      
 {\usepackage[utf8]{inputenc}}           

\usepackage[USenglish]{babel}

\ifboolexpr{bool{jacowbiblatex}}%
 {%
  \addbibresource{WEP085.bib}
  \addbibresource{biblatex-examples.bib}
 }{}
\begin{document}

\title{Start-to-End Simulations of a Compact, Linac-Based Positron Source}

\author{S. Crisp \thanks{scrisp11@slac.stanford.edu}, R. Goldman\thanks{also at University of California, Los Angeles}, A. Ismail\thanks{also at Stanford University}, S. Gessner\thanks{also at Stanford University}, \\ SLAC National Accelerator Laboratory, Menlo Park, USA
		}
	
\maketitle

\begin{abstract}
Slow positrons are increasingly important to the study of material surfaces. For these kinds of studies, the positrons must have low emittance and relatively high brightness. Unfortunately, fast positron sources like radioactive capsules or linac driven sources have broad energy and angular spread, which make them difficult to capture and use. Moderators are materials that produce slow, mono-energetic positrons from a fast positron beam. Since their efficiencies are typically less than $10^{-3}$ slow $e^+$ per fast $e^+$, research into how to maximize efficiency is of great interest. Previous work has shown that using a linac, one can decelerate the fast positron beam in order to greatly increase moderation efficiency. We present here start-to-end simulations using G4beamline to model a 100~MeV electron beam incident upon a Tungsten target, focused by an adiabatic matching device, and decelerated by a 1.3~GHz, 5-cell pillbox cavity. We show that by decelerating the positrons after their creation we can increase the number of positrons under 500~keV by 15 times, translating to a 16.3 times improvement in moderation efficiency, and therefore leading to a brighter positron source.
\end{abstract}

\section{Introduction}
Slow positron facilities around the world enable basic science through the use of properties unique to positrons. At the NEPOMUC positron facility, an intense positron source has been employed to drive positron annihilation lifetime spectroscopy to measure vacancy-like defects through the use of positron lifetime studies \cite{Egger_2010_PLEPS}, as well as a multitude of other instruments for surface and volume characterization. At KEK, Total Reflection High-Energy Positron Diffraction (TRHEPD) takes advantage of the unique ability of positrons to totally reflect from surfaces to measure only a few atomic layers deep \cite{Hyodo_2014_TRHEPD}. Doing this quickly and with high resolution requires an intense (>$10^9$ $e^+$/s) slow positron beam, where slow is generally defined as less than $30-40$~keV \cite{HUGENSCHMIDT_2016547_posreview}. At SLAC National Accelerator Laboratory, we are developing a Compact Positron Source~\cite{Hessami2023} and are interested in producing beams which are both high intensity and ultrashort to complement the existing MeV-UED facility \cite{Weathersby2015}. This would add yet another tool to the positron physics toolkit -- Ultrafast Positron Diffraction, which would enable an entirely new area of surface sensitive pump-probe experiments. This kind of experiment requires maximizing the efficiency of the positron source, since without large numbers of particles these measurements would be too time-intensive to take.

Fast positrons are initially generated either by use of $\beta^+$ emitters like $^{22}$Na or through pair production, which can provide a much higher intensity source. In either case, these fast positrons have high divergence and energy spread, so while one could simply capture a particular energy slice, this excludes the use of the vast majority of the initial positrons. Slow positron facilities instead use a process called moderation, which cools the positron distribution into a nearly mono-energetic beam, with the energy distribution essentially being a thermal spectrum \cite{Fischer_1986_ThermalModerator}. 

These slow positrons have significantly reduced energy spread and emittance such that despite moderators having efficiencies of $\epsilon\approx10^{-3}$ slow positrons per incident positron, the beams which exit a positron moderator are several orders of magnitude brighter. Although rare gas solid moderators like  neon have been demonstrated to have $\epsilon\approx10^{-2}$, these degrade quickly in high flux environments, such as in linac based positron sources. In the simplest case, a moderator is a thin metal foil for which positrons have a negative work function. This means that a positron which scatters throughout the fields and thermalizes before annihilation has some probability of diffusing to the surface and being reemitted with energy equal to $\phi$ ($\phi_W = -3$~eV), and energy spread equal to the thermal energy, on the order of meV \cite{coleman2000positronBook}. Some facilities follow this initial moderator by a remoderation stage \cite{Mills1980_Reomoderation}, or by the use of a buffer gas trap which can both accumulate and further cool positron beams \cite{Surko_1989_buffergastrap}, both of which further increase beam brightness. For this paper, we will limit ourselves to the use of a single $50$~$\mu$m Tungsten foil as a moderator, while acknowledging that further work is necessary to fully optimize the moderator geometry.

Whilst other slow positron facilities utilize various complexities of target-moderator assemblies, none so far has experimentally demonstrated an increase moderation efficiency by preemptively slowing the positron bunch before the initial moderation stage. At CLIC, the positron bunch is slowed in the cell of the linac, but this is used only to capture more charge in a bucket before accelerating to higher energy for collider applications \cite{Han_2019_CLICoptRF}. In this paper, we follow on the work of Long et al.~\cite{Long_2007_decelLinac}, who found via simulation that using a 108 MHz cavity tuned to a decelerating phase between the target and moderator could increase the moderator efficiency by 20 times. In this paper, we present a design based on a 100~MeV electron accelerator source combined with a 1.3~GHz cavity to imitate an already fabricated cavity at SLAC.

\section{Simulation}

\begin{figure}
    \centering
    \includegraphics[width=\linewidth]{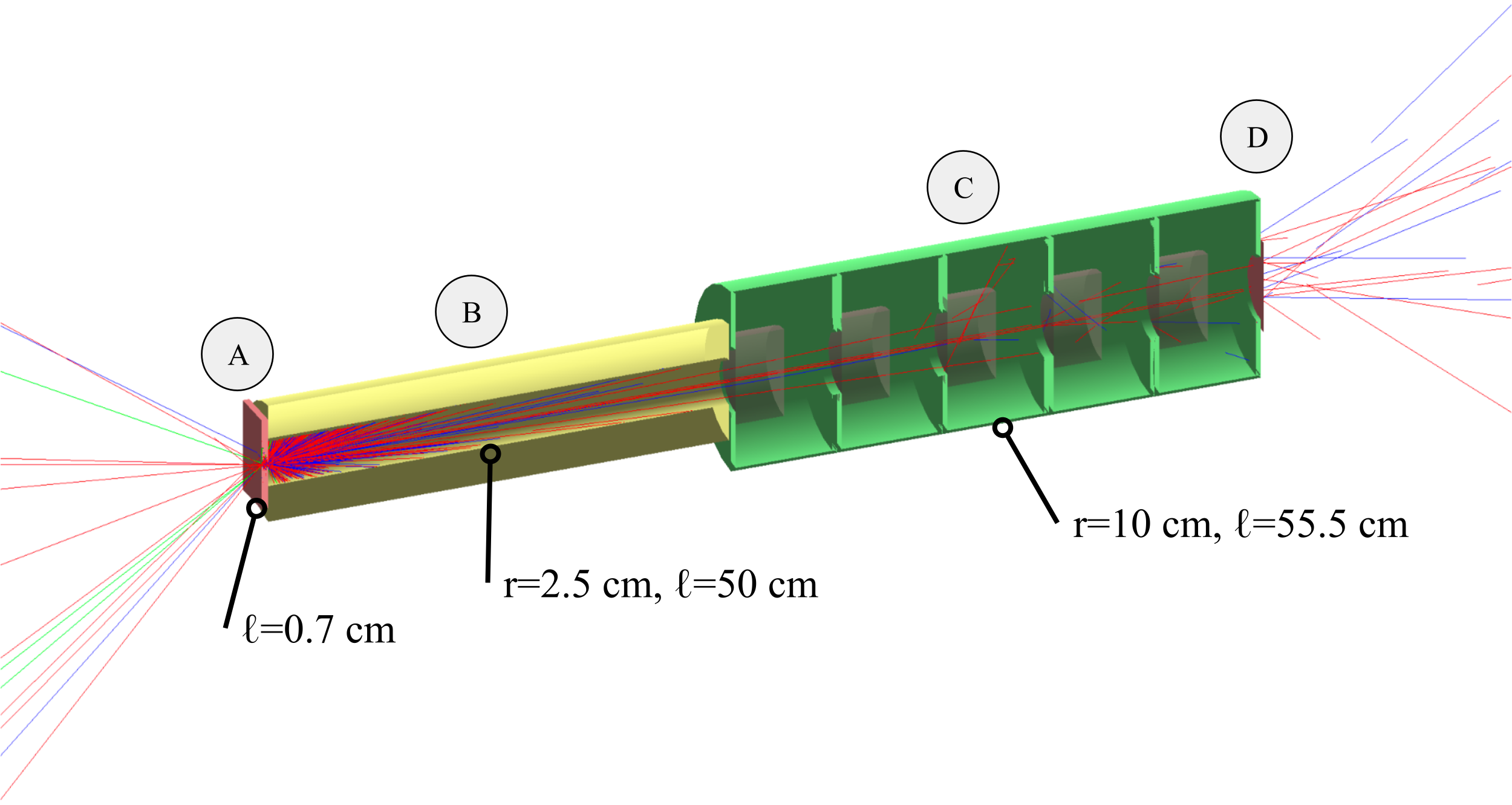}
    \caption{G4Beamline simulation setup. Red traces corresponded to electrons, blue to positrons, and green to gamma rays. Gamma rays after the target are not shown, for easier viewing. The electron beam of 100~MeV energy first hits the (a) Tungsten target, from which fast positrons are extracted into the (b) Adiabatic Matching Device capture section. After this, the positrons are slowed by the (c) L-band RF stage. The positrons then hit a 50~micron (d) Tungsten moderator from which we extract the stopping profile.}
    \label{fig:schematic}
\end{figure}

The particle tracking and simulation code G4beamline, based on the Geant4 toolkit \cite{agostinelli2003geant4}, is used to simulate the entire beamline from start-to-end, pictured in Fig. \ref{fig:schematic}. In order to calculate positron stopping profiles within the Tungsten moderator, care must be taken to use a physics model which sufficiently describes sub-keV physics. In accordance with other sources, we use the Penelope-2008 as implemented in Geant4, which can model physics down to 100~eV \cite{orourke_2011_simulations}. Because the simulation cannot model the physics of thermalization and diffusion within the moderator, we instead apply an energy cutoff of 50~eV, at which we assume that the positron has a probability of diffusion back to the surface based on its position, which does not change substantially in the relevant timescale.

\begin{figure}
    \centering
    \includegraphics[width=\linewidth]{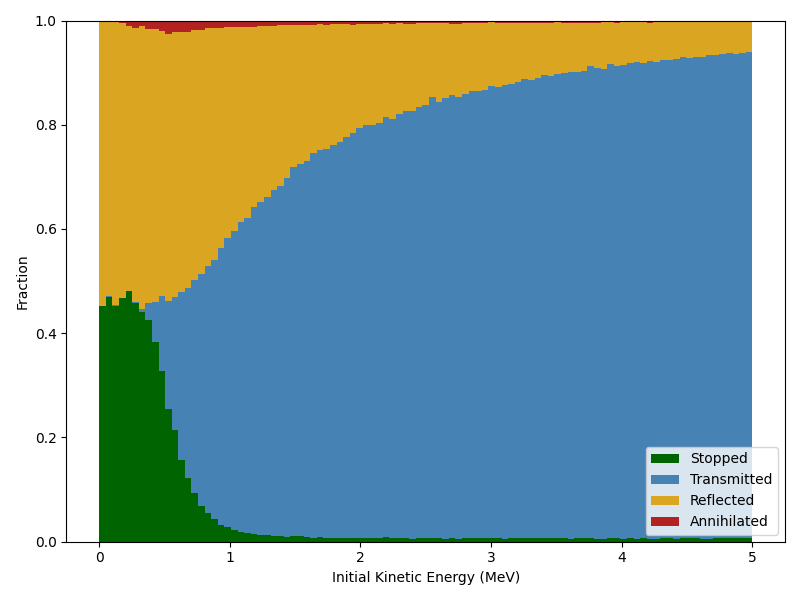}
    \caption{A graph showing the various end states for a positron incident on a 50~$\mu$m W foil acting as a moderator. In order to be moderated, a positron must be stopped within a small distance from the moderator surface. Few particles with energies greater than 1~MeV contribute to the moderated positron population. For energies less than 0.5~MeV, 41.5\% are stopped.}
    \label{fig:moderatorEfficiency}
\end{figure}

In Fig. \ref{fig:moderatorEfficiency}, we show that this moderation process is extremely energy dependent. We model the moderation by simulating single-positrons with specified kinetic energy ($E_k$) incident onto a $50$~$\mu$m W foil. Detectors are placed before and after the moderator. We see that beyond 1~MeV, positrons are unlikely to stop within the foil, instead simply scattering in the material and flying through. Since only the particles which stop within the moderator contribute the the slow positron beam, we must maximize the number of fast positrons with energy less than 1~MeV. It should be noted that the distance into the moderator at which the positrons stop is also extremely important to deciphering moderator efficiency. Further work will investigate this in order to maximize total moderator efficiency. As a rule of thumb, the higher the energy of the fast positron, the deeper into the moderator it will propagate. Only those positrons which stop very near (sub-micron) to the edge of the moderator will be able to escape the material.

Figure~\ref{fig:schematic} shows the beamline simulated in G4Beamline. 
We use a 100~MeV drive $e^-$ beam, with parameters consistent with the XTA beamline at SLAC \cite{Limborg2012_XTA,Morton2025_NLCTA}. The pulse length of the electron drive bunch is on the order of 100~fs. The outgoing positron shower is longer in time because the positrons have large energy spread, so the initial time distribution of the electrons is not relevant for the simulation. The target is a 6.5~mm, or 1.86 radiation length, thick piece of tungsten, optimized to maximize the number of fast positrons produced; we obtain 0.23~$e^+$/$e^-$ for the 100~MeV drive.

Because the resulting particle shower has a large angular spread, we require strong focusing in order to confine particles into the linac. To do this, we use an adiabatic matching device (AMD)~\cite{Chehab_1983_OrsayAMD}, which serves to rotate the outgoing phase space from the target. That is, the AMD turns a small spot size and large transverse momentum, $p_T$, into a large transverse spot size with small $p_T$. The AMD is defined by the field 

\begin{equation}
    B_z = \frac{B_0}{1 + \alpha z},
\end{equation}

Where $B_0$ is generally the maximum attainable field and $\alpha$ is the parameter which defines the steepness of the field. If we define that at the length $L$ of the AMD, we reach a magnetic field of $B_s$, then we can find the transverse momentum acceptance of such a system is 

\begin{equation}
    p_T = e\sqrt{B_0B_s}a
\end{equation}

where $e$ is the charge of the positron and $a$ is the aperture radius \cite{Chehab_1983_OrsayAMD}.

Here, we use a peak magnetic field $B_0 = 5$~T which decays to $B_s = 0.5$~T over 0.5~m, corresponding to $\alpha = 18\text{ m}^{-1}$. Combined with a cylindrical aperture of 2.5~cm, we have a momentum acceptance of up to 11.8~MeV/c. With this AMD, we are able to collect 69.9\% of the fast positrons. The temporal length of the positron beam is extended from $\sigma_t=11.9$~ps at the start of the AMD to $\sigma_t=110.3$~ps at the end. This positron pulse length is proportional to the length of the AMD, so further exploration of AMD parameters is necessary to fully optimize the system for injection into the cavity and maximize the final beam brightness. 

We model the decelerating cavity as a series of pillboxes with an aperture radius of $3$~cm and operating frequency of $1.3$~GHz. We assume a constant $B_z = 0.5$~T field throughout the cavity, which could be produced by a conventional solenoid. This mirrors the cavity originally designed for the ILC, which has been tested at SLAC up to a gradient of 13.7~MV/m \cite{Wang2009_lbandILC}. To optimize, the phase of each cavity is treated as independent and we use the Monte Carlo simulated annealing algorithm as the search method for optimal phases. 

\begin{figure}
    \centering
    \includegraphics[width=\linewidth]{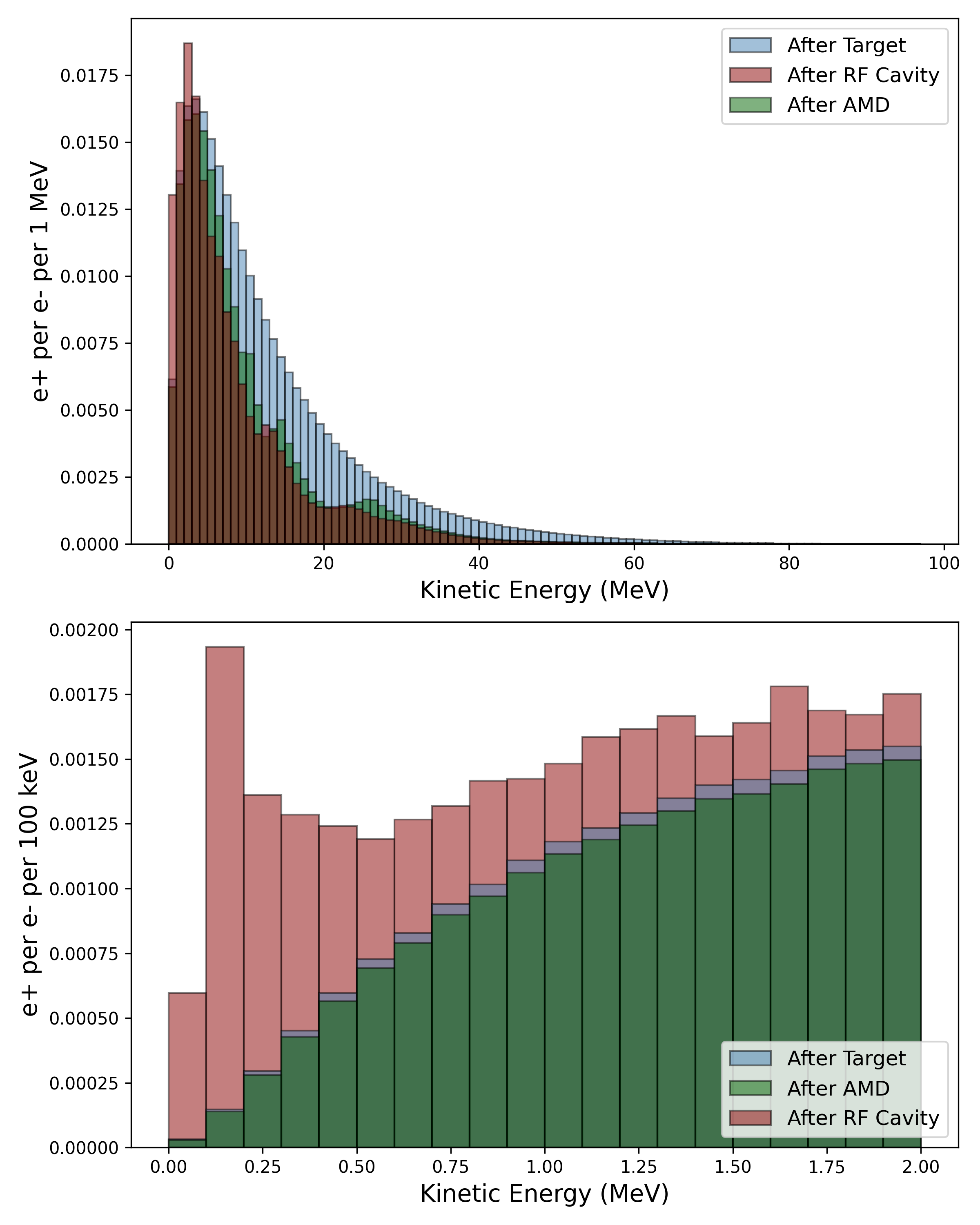}
    \caption{Kinetic energy ($E_k$) histograms directly after the target, at the output of the AMD, and after the linac. At top, we see that positrons directly after the target have broad energy spread. In the AMD section we are able to confine 69.9\% of particles, with most losses being from $E_k>10$~MeV positrons. On bottom, we zoom in on positrons with $E_k<1$~MeV. After optimizing the L-band cavity we have 4.5 times more positrons with $E_k<0.5$~MeV.}
    \label{fig:linacKE_Hist}
\end{figure}

Figure \ref{fig:linacKE_Hist} shows that through the use of this optimization we can decrease the particle energy such that 15 times more positrons have $E_k<200~$keV compared to the spectrum directly after the target. In total, there is a 16.3-fold increase in stopped positrons in the 50~$\mu$m Tungsten foil moderator by using the L-band cavity. This is at the cost of bunch length -- at the end of the linac, the positron beam has spread out across $\approx5$~ns, although $\sigma_t=355$~ps for particles with $E_k<500$~keV.

\section{Conclusion}
In this paper, we have outlined the current status of the design process for a linac based positron source which takes advantage of the energy sensitivity of the moderation process. Through the use of a decelerating L-band cavity, we are able to increase the number of stopped positrons in a positron moderator by 16.3~times. Future work will explore the use of buffer gas traps to further increase brightness as well as bunching techniques and emittance minimization. We plan to test these concepts at the XTA beamline at SLAC.

\section{Acknowledgments}
This work was supported by the Department of Energy, Laboratory Directed Research and Development program at SLAC National Accelerator Laboratory, under contract DE-AC02-76SF00515. This work was supported in part by the U.S. Department of Energy, Office of Science, Office of Workforce Development for Teachers and Scientists (WDTS) under the Science Undergraduate Laboratory Internships (SULI) program.
%
\ifboolexpr{bool{jacowbiblatex}}%
	{\printbibliography}%
	{%
	
	
} 
%
%


\end{document}